\input harvmac

\def\figin{\epsfcheck\figin}\def\figins{\epsfcheck\figins}
\def\epsfcheck{\ifx\epsfbox\UnDeFiNeD
\message{(NO epsf.tex, FIGURES WILL BE IGNORED)}
\gdef\figin##1{\vskip2in}\gdef\figins##1{\hskip.5in}
instead
\else\message{(FIGURES WILL BE INCLUDED)}%
\gdef\figin##1{##1}\gdef\figins##1{##1}\fi}
\def\DefWarn#1{}
\def\figinsert{\goodbreak\midinsert}
\def\ifig#1#2#3{\DefWarn#1\xdef#1{fig.~\the\figno}
\writedef{#1\leftbracket fig.\noexpand~\the\figno}%
\figinsert\figin{\centerline{#3}}\medskip\centerline{\vbox{\baselineskip12pt
\advance\hsize by -1truein\noindent\footnotefont{\bf Fig.~\the\figno:}
#2}}
\bigskip\endinsert\global\advance\figno by1}

\overfullrule=0mm

\def\IR{\relax{\rm I\kern-.18em R}}
\font\cmss=cmss10 \font\cmsss=cmss10 at 7pt
\def\IZ{\relax\ifmmode\mathchoice
{\hbox{\cmss Z\kern-.4em Z}}{\hbox{\cmss Z\kern-.4em Z}}
{\lower.9pt\hbox{\cmsss Z\kern-.4em Z}}
{\lower1.2pt\hbox{\cmsss Z\kern-.4em Z}}\else{\cmss Z\kern-.4em Z}\fi}

\def\inbar{\,\vrule height1.5ex width.4pt depth0pt}
\def\IB{\relax{\rm I\kern-.18em B}}
\def\IC{\relax\hbox{$\inbar\kern-.3em{\rm C}$}}
\def\ID{\relax{\rm I\kern-.18em D}}
\def\IE{\relax{\rm I\kern-.18em E}}
\def\IF{\relax{\rm I\kern-.18em F}}
\def\IG{\relax\hbox{$\inbar\kern-.3em{\rm G}$}}
\def\IH{\relax{\rm I\kern-.18em H}}
\def\II{\relax{\rm I\kern-.18em I}}
\def\IK{\relax{\rm I\kern-.18em K}}
\def\IL{\relax{\rm I\kern-.18em L}}
\def\IM{\relax{\rm I\kern-.18em M}}
\def\IN{\relax{\rm I\kern-.18em N}}
\def\IO{\relax\hbox{$\inbar\kern-.3em{\rm O}$}}
\def\IP{\relax{\rm I\kern-.18em P}}
\def\IQ{\relax\hbox{$\inbar\kern-.3em{\rm Q}$}}
\def\IGa{\relax\hbox{${\rm I}\kern-.18em\Gamma$}}
\def\IPi{\relax\hbox{${\rm I}\kern-.18em\Pi$}}
\def\ITh{\relax\hbox{$\inbar\kern-.3em\Theta$}}
\def\IOm{\relax\hbox{$\inbar\kern-3.00pt\Omega$}}


\def\~{\tilde }
\def\^{\hat }
\def\={\bar }

\def\bbuildrel#1_#2^#3{\mathrel{\mathop{\kern 0pt#1}\limits_{#2}^{#3}}}

\catcode`\@=11
\def\displaylinesno#1{\displ@y\halign{
	\hbox to\displaywidth{$\@lign\hfil\displaystyle##\hfil$}&
	\llap{$##$}\crcr
#1\crcr}}
\catcode`\@=12

\def\sitle#1#2{\nopagenumbers\abstractfont\line{#1}%
\vskip .5in\centerline{\titlefont #2}\abstractfont\vskip .2in\pageno=0}

\def\myfnii{e-mail: sochen@asterix.lbl.gov }


\def\acknowledge{
This work was supported in part by the Director, Office of Energy
Research, Office of High Energy and Nuclear Physics, Division of High
Energy Physics of the U.S. Department of Energy under Contract
DE-AC03-76SF00098 and in part by the National Science Foundation  
under
grant PHY-90-21139.
}



\sitle{  \hfill LBL-38034, UCB-PTH-96/04}
{Integrable Generalized Principal Chiral Models}
\centerline{ Nir Sochen\footnote{$^*$}{\myfnii} }
\vskip0.4cm
\centerline{\it Department of Physics\footnote{$^\dagger$}{\acknowledge}}
\centerline{\it University of California at Berkeley }
\centerline{\it and} 
\centerline{\it Theoretical Physics group} 
\centerline{\it Lawrence Berkeley Laboratory} 
\centerline{\it Berkeley, CA 94720, U.S.A.} 
\vskip0.4cm
\centerline{\bf Abstract }
We study 2D non-linear sigma models on a group manifold with 
a special form
of the metric. We address the question of integrability for this special 
class of sigma models. We derive two algebraic conditions for the metric on 
the group manifold. Each solution of these conditions defines an integrable 
model. Although the algebraic system is overdetermined in general, we give
two examples of solutions. We find the Lax field for these models and calculate
their Poisson brackets. We also obtain the renormalization group (RG)
equations, to first order, for the generic model. We solve the RG equations
for the examples we have and show that they are integrable 
along the RG flow. 

\Date{Dec. 95}
\def\buildrel#1\over#2{\mathrel{
     \mathop{\kern 0pt#2}\limits^{#1}}}
\def\half{{1\over 2}}
\def\cM{{\cal M}}
\def\cL{{\cal L}}
\def\lab{L_{ab} }
\newsec{Introduction}
A lot of efforts have been dedicated in the last decade to the 
determination, classification and duality relations between 2D non-linear 
sigma models who are conformal. Since the $\beta$ function of the sigma 
model appears in a geometrical form 
\ref\DF{D. Friedan, (1980) ``Non linear models in 2+$\epsilon$ dimensions",
Ph.D. thesis, published in Ann. Phys. 163 (1985) 318. } 
and since conformal invariance manifests itself by the
vanishing of the $\beta$ function, we end up having a geometrical 
criterion for the class of conformally invariant non-linear sigma models. 

What can
be said about non-linear sigma models if we change the condition of conformal 
invariance to the more general condition of integrability?

In this paper we take a first step toward an answer to this question.
We are limiting ourselves to Lie group manifolds and taking a special
form for the metric. This class, that we call generalized principal chiral 
model (GPCM), is a generalization of the diagonal anisotropic
principal chiral model 
\ref\KR{I.V. Cherednik, Theor. Math. Phys. 47 (1981) 225.\semi
See also\semi
A.N. Kirilov and N. Yu. Reshetikhin, Paris 1986, Proceedings, ``String Theory,
Quantum Cosmology and Quantum Gravity: Integrable and Conformal Invariant
Theories" pp. 235-257, 
and references therein.}.
It is also equivalent to the generalized Thirring
model studied recently by Bardak\c ci and Bernardo 
\ref\BB{K. Bardak\c ci and L. Bernardo, Nucl. Phys. B450 (1995) 695.}.
While they were looking 
for conditions on the coupling constants to insure conformal invariance,
we will derive, for this class of models,  conditions
on the metric (or coupling constants) to insure classical integrability. This
is done by demanding that a Lax pair formulation for the equation of motion is
possible. Once we have the Lax field and the Poisson structure on the phase
space, we calculate the Poisson brackets. We find a 
surprise in the PCM model because a set of operators, which includes the Lax
field, not only closes under the Poisson brackets but also forms an algebra,
very similar to the two loop affine Lie algebra 
\ref\Schwim{L.A. Ferreira,J.F. Gomes,
A.H Zimermman and A. Schwimmer, Phys. Lett. B274 (1992) 65.}.

Since integrable models are not, in general, at the fixed point of the 
renormalization group (RG) equations, it is interesting to know whether 
they remain integrable along the RG flow. In this paper 
we find the $\beta $ function, to first order, for the generic model
in the class that we consider, and solve the RG equations for two examples.
For these two cases the models are integrable along the RG flow.

\newsec{Equations of motion}

The models that we consider are generalizations of the principal chiral model
(PCM) defined on a Lie group manifold $G$.
We replace the trace over the algebra by a more general bilinear form.
The action is
\eqn\eqact{
I=-\int d\sigma d\tau L_{ab}(g^{-1}\del_\mu g)^a (g^{-1}\del^\mu g)^b
}
where $g\in G$ and $A_\mu A^\mu=A_\mu A_\nu \eta^{\mu\nu}$ with 
$\eta^{00}=-\eta^{11}=1$, $\eta^{01}=\eta^{10}=0$. Here, $x_0\equiv\tau$, 
$x_1\equiv\sigma$, and $\lab $ is taken to be a symmetric and invertible 
$\dim G\times \dim G$ matrix. 

We introduce a flat connection by the standard definition:
\eqn\eqcon{
A_\mu=-ig^{-1}\del_\mu g=A_\mu^at_a
}
where $t_a$ are the generators of the Lie algebra ${\cal G}=$Lie($G$):
\eqn\Liecom{
[t_a,t_b]=if_{ab}^{\ \ c}t_c\quad .
}
The metric on the group manifold is given by $\kappa_{ab}=Tr(t_at_b)$.
The inverse is denoted by $\kappa^{ab}$ such that 
$
\kappa^{ac}\kappa_{cb}= \kappa_{bc}\kappa^{ca}= \delta^a_b
$.
The Bianchi identity for the connection reads
\eqn\eqBian{
\del_\mu A_\nu^a-\del_\nu A_\mu^a-f_{bc}^{\ \ a} A_\mu^bA_\nu^c=0\quad .
}

Next we derive the equations of motion. Define $\delta\rho=-ig^{-1}\delta g$,  
then
\eqn\eqdelA{
\delta A_\nu^a=\del_\nu\delta\rho^a-f_{bc}^{\ \ a}A_\nu^b\delta\rho^c
}
and
\eqn\eqdelI{
-{\delta I\over\delta\rho^c}=L_{ac}\del_\mu A^{\mu a}+
\lab f_{dc}^{\ \ b} A_\mu^a A^{\mu d}=0\quad .
}
Since $L$ is invertible we can write the equations of motion in the form
\eqn\eqmot{
d_\mu A^{\mu a}=0
}
where
\eqn\eqFabc{
\eqalign{
 d_{\mu b}^a&=\delta^a_b\del_\mu+S^a_{\ bc}A^c_\mu\cr
 S_{\ bc}^a &= \half (F_{\ bc}^a  + F_{\ cb}^a )\cr
 F_{\ bc}^a &= L_{qb}f_{cp}^{\ \ q}(L^{-1})^{pa}\quad .\cr
}
}
We see that $S_{\ bc}^a$ is playing a role of a 
connection. This will also be apparent below when we discuss the general
non-linear sigma model on a group manifold.

\newsec{The Lax pair formulation}

A basic theorem in classical field theory 
asserts that a model is integrable if
its equation of motion can be represented as a one parametric Lax pair:

\eqn\eqLax{
[\del_0+\cM(\lambda),\del_1+\cL(\lambda)]=0
}
where $\lambda$ is the spectral parameter.
We will now derive conditions on $\lab$ such that a Lax pair representation
is possible. 

Because the equations of motion are quadratic in the currents, we take as
an ansatz a commutator between linear combination of the currents:
\eqn\eqansatz{
[\del_0 + N_{0b}^aA_0^bt_a+N_{1b}^aA_1^bt_a,
\del_1 + N_{0b}^aA_1^bt_a+N_{1b}^aA_0^bt_a]=0
}
where $N_{\mu b}^a$ are two unknown auxiliary $\dim G\times\dim G$ matrices.
This linear structure is also present in known integrable models of this type.

Computing explicitly the commutator, using \Liecom\ and the Bianchi identity,
we obtain an equation
\eqn\eqexplic{
\big(N^s_{0b}f_{pq}^{\ \ b}+
iN^a_{\mu p}N^{\mu c}_{\ \ q}f_{ac}^{\ \ s}\big)A^p_0A^q_1
+N^s_{1b}\del_\mu A^{\mu b}+
iN^a_{0p}N^c_{1q}f_{ac}^{\ \ s}A_\mu^pA^{\mu q}=0
}
which gives rise upon comparison with the equations of motion to the 
following conditions:
\eqna\eqmaster
$$
\displaylinesno{
N^a_{0b}f_{pq}^{\ \ b}+
iN^b_{\mu p}N^{\mu c}_{\ \ q}f_{bc}^{\ \ a}=0&\eqmaster a\cr
\half if_{cd}^{\ \ a}\big(N^c_{0p}N^d_{1q}+N^c_{0q}N^d_{1p}\big)=
N^a_{1b}S_{\ pq}^b &\eqmaster b\cr
}
$$
The second equation relates the ansatz parameters to the coupling constant
metric $\lab$ by eq. \eqFabc.

The solution to these two algebraic equations is an affine variety $\cM$.
The model is integrable if ${\dim \cM} > 0$. A naive counting shows 
that we have $2(\dim G)^3$ equations for only $2(\dim G)^2$ variables.
A generic choice of a matrix $\lab$ is, as expected, not integrable.
Nevertheless the set of solutions of \eqmaster{}\ is not empty.
 
Before proceeding to the examples we would like to comment on the 
generalization
where $L_{ab}$ is not a constant function on the group manifold.
In this case the auxiliary matrices $N_{\mu b}^a$ vary
on the group manifold. Nevertheless the conditions \eqmaster{}\
still hold with the only change 
$S_{\ pq}^a\to \tilde S_{\ pq}^a= S_{\ pq}^a+ \Gamma_{\ pq}^a$
where $\Gamma^a_{\ pq}$ is a connection
\eqn\eqgam{
\Gamma_{\ pq}^a
=\half (L^{-1})^{ab}\big(
\del_p L_{bq}+
\del_q L_{bp}-
\del_b L_{pq}\big)\quad .
}
with the notation $\del_a\equiv e_a^i\del_i$.
Note that $S^a_{\ pq}$ is playing a role of a connection.

We also have, in this case, chirality conditions on $N_\mu$:
\eqn\Npm{
\del_+N_-=\del_-N_+=0
}
where $\del_\pm=\del_0\pm\del_1$ and similarly $N_\pm=N_0\pm N_1$.
In a light cone formulation the conditions \eqmaster{}\ read:
\eqn\eqmaspm{
\eqalign{
(N_++N_-)^s_bf_{pq}^b+i(N_{+p}^aN_{-q}^c+N_{-p}^aN_{+q}^c)f_{ac}^s&=0\cr
(N_+-N_-)^s_b\tilde S_{pq}^b
+i\half (N_{+p}^aN_{-q}^c-N_{-p}^aN_{+q}^c)f_{ac}^s&=0\cr
}
}
For a detailed study of the general integrable non linear sigma model on 
a group manifold see 
\ref\NSI{N. Sochen, ``Integrable sigma models on a group manifold", 
In preparation}.

Next we show how some known models fit into this framework.

$\underline{\hbox{Example 1}}$: The principal chiral model (PCM) where 
$\lab={1\over g^2}\kappa_{ab}$ is known
to be integrable \ref\PW{A. M. Polyakov and P. B. Wiegmann. Phys. Lett. 131B 
(1984) 121\semi
P. B. Wiegmann, Phys. Lett. 141B (1984) 217\semi
P. B. Wiegmann, Phys. Lett. 142B (1984) 173\semi
E. Ogievetski, N. Yu. Reshetikhin and P. B. Wiegmann, Nucl. Phys. B280 [FS18]
 (1987) 45.}
and upon taking $N^a_{\mu b}=\lambda_\mu\delta^a_{b}$ we
find that the system \eqmaster{}\ is reduced to one equation for two variables
\eqn\eqpcm{
\lambda_0+i(\lambda_0^2-\lambda_1^2)=0
}
we can write the solution in a parametric way
\eqn\wqlam{
\eqalign{
\lambda_0&=-i\sinh^2\lambda\cr
\lambda_1&=-i\sinh\lambda\cosh\lambda\cr
}
}

$\underline{\hbox{Example 2}}$: The diagonal anisotropic $SU(2)$ PCM with 
$\lab=J_a\kappa_{ab}$ 
(no summation) with $J_1=J_2\ne J_3$ is also a known integrable model \KR.
Taking $N_{\mu b}^a = \lambda_\mu^a\delta^a_b$ (no summation) and
assuming $\lambda_\mu^1=\lambda_\mu^2\ne\lambda_\mu^3$ gives 
three equations 
\eqn\equnisopcm{
\eqalign{
\lambda_0^1+i(\lambda_0^1\lambda_0^3-\lambda_1^1\lambda_1^3)&=0\cr
\lambda_0^3+i((\lambda_0^1)^2-(\lambda_1^1)^2)&=0\cr
\lambda_0^1\lambda_1^3-\lambda_1^1\lambda_0^3&=
{J_1-J_3\over J_1}\lambda_1^1\cr
}
}
for four variables: $(\lambda_\mu^1,\lambda_\mu^3),\quad \mu=0,1$.
In parametric representation it reads
\eqn\eqparam{
\eqalign{
\lambda^1_1&={1\over m}\sinh\lambda\cosh\lambda\cr
\lambda^1_0&={1\over m}\sinh\lambda(m^2+\cosh^2\lambda)^{1\over 2}\cr
\lambda^1_3&=-i\cosh\lambda(m^2+\cosh^2\lambda)^{1\over 2}\cr
\lambda^0_3&=-i\sinh^2\lambda\cr
}
}
where $k={J_1-J_3\over J_1}$ and $m^2=ik-1$.

More analysis is needed in order to find new non trivial solutions. Work
in this direction is in progress and will be reported elsewhere.

\newsec{The fundamental Poisson brackets}

We take in this section a canonical approach and use a technique
first used by Bowcock
\ref\Bo{P. Bowcock, Nucl. Phys. B316 (1989) 80.}.
In order to compute the Poisson brackets $\{\cL^a(\sigma),\cL^b(\sigma ')\}$ it
is convenient to choose a local coordinate system $x^i$. In terms of these 
coordinates we write
\eqn\eqAmu{
A^a_\mu=-i(g^{-1}\del_\mu g)^a=e_i^a(x)\del_\mu x^i
}
where
\eqn\eqeai{
e^a_i(x)=-i\big(g(x)^{-1}\del_i g(x)\big)^a\quad 
}
are the vielbeins.

The action now is in the general form of a non linear sigma model
\eqn\sigform{
I=\int d\tau d\sigma G_{ij}(x)\del_\mu x^i\del^\mu x^j
}
where the metric $G_{ij}(x)$ is given 
\eqn\eqGij{
G_{ij}(x)= L_{ab}e_i^a(x)e_j^b(x)
}
in terms of the vielbeins.

The canonical conjugate momentum is
\eqn\conjpi{
\pi_i(\sigma)={\delta I\over\delta {\dot x}^i}=2G_{ij}{\dot x}^j
} 
where ${\dot x}^i={\del\over \del\tau} x^i$ and we will also use below the
notation $f'={\del\over \del\sigma} f$.
The canonical bracket is
\eqn\canbra{
\{x^i(\sigma),\pi_j(\sigma ')\}=\delta^i_j\delta(\sigma -\sigma ')
}
The ``gauge potentials'' $A^a_\mu$ are functions on the phase space given by
\eqn\eqAphase{
\eqalign{
A^a_0&=e^a_i{\dot x}^i=\half (L^{-1})^{ab}e_b^i\pi_i\cr
A^a_1&=e^a_i{x'}^i\cr
}
}
where $e_b^i$ 
\eqn\eqeinv{
e_i^be_b^j=\delta_i^j\quad ; \quad e_i^ae_b^i=\delta_b^a
}
is the inverse of $e_i^b$.

Using the standard identities
\eqna\eqiden
$$
\displaylinesno{
\del_ie^a_j-\del_je^a_i=e_i^be_j^cf_{bc}^{\ \ a}&\eqiden a\cr
f(\sigma ')\delta '(\sigma-\sigma ')-
f(\sigma)\delta '(\sigma-\sigma ')=
f'(\sigma)\delta(\sigma-\sigma ')&\eqiden b\cr
{\del\over\del\sigma '}\delta(\sigma-\sigma ')=
-{\del\over\del\sigma}\delta(\sigma-\sigma ')&\eqiden c\cr
}
$$
one can obtain the commutation relations
\eqn\eqAmuAnu{
\eqalign{
\{\tilde A^c_0(\sigma),A^d_1(\sigma ')\}&=f_p^{\ cd}A_1^p
\delta(\sigma -\sigma ')
+\kappa^{cd}\delta '(\sigma -\sigma ')\cr
\{\tilde A^c_0(\sigma),\tilde A^d_0(\sigma ')\}&=f_p^{\ cd}\tilde A_0^p
\delta(\sigma -\sigma ')\cr
}
}
where 
\eqn\atil{
\tilde A_0^a=2\kappa^{ab}L_{bc}A_0^c\quad .
}

These commutation relations
are the building blocks for the calculation of the Poisson bracket for
the Lax field. Recall the definition
\eqn\eqLdef{ 
\cL^a(\sigma)= N_{0b}^aA_1^b+N_{1b}^aA_0^b.
}
With the help of \eqAmuAnu, a straightforward calculation gives
\eqn\eqLaLb{
\{\cL^a(\sigma),\cL^b(\sigma')\}=\half (\Gamma^{ab}_{0q}A_1^q
+ \Gamma^{ab}_{1q}A_0^q)
\delta(\sigma -\sigma ')
+\half Q^{ab}
\delta '(\sigma -\sigma ')
}
where
\eqn\eqdef{
\eqalign{
\Gamma^{ab}_{0q}&=
(N^a_{0c}N^b_{1d}-N^b_{0c}N^a_{1d})(L^{-1})^{ds}f_{sq}^{\ \ c}\cr
\Gamma^{ab}_{1q}&=
N^a_{1c}N^b_{1d}(L^{-1})^{cs}(L^{-1})^{dr}f_{sr}^{\ \ p}L_{pq}\cr
Q^{ab}&=(N^a_{0c}N^b_{1d}+N^b_{0d}N^a_{1c})(L^{-1})^{dc}.\cr
}
}
This computation did not use the integrability conditions which relate
the auxiliary $N$ matrices with the coupling constants matrix $L$. Therefore we
cannot expect the Poisson bracket to close simply on some algebra. In
fact even using \eqmaster{}\ does not simplify things, and we were not
able to identify, for the generic case, a set of fields which include 
the Lax field and is closed under the Poisson bracket.
Although the Poisson bracket $\{\cL^a(\sigma),\cL^b(\sigma')\}$ 
at equal $\tau$ 
does not close in general, for the case of the PCM 
(i.e. $L_{ab}={1\over g^2}\kappa_{ab}$)
we have the following:
\eqn\eqLPCM{
\cL^a=\lambda_0A_1^a+\lambda_1A_0^a=-i\sinh(\lambda )
(\sinh(\lambda) A_1^a+\cosh(\lambda) A_0^a)
}
Let us define the following fields
\eqn\eqLnPCM{
J^a_n={2\over g^2}(n\tanh(\lambda) A_1^a+A_0^a)
}
where $n\in Z$, and our Lax field is proportional to $J^a_1$:
\eqn\eqlaz{
\cL^a=-{i\over 4}g^2\sinh 2(\lambda) J^a_1\quad .
}
These operators generate an algebra very similar to the two loop affine
algebra.
\eqn\eqLaLbPCM{
\{J^a_n(\sigma),J^b_m(\sigma')\}=
f^{ab}_{\ \ c}J^c_{n+m}(\sigma)\delta(\sigma-\sigma')+
{2(n+m)\over g^2}\tanh(\lambda)\kappa^{ab}\delta '(\sigma-\sigma ')
}
from which it is now straightforward to find the fundamental Poisson brackets:
\eqn\eqLanLbm{
\{J_n(\sigma){\buildrel\otimes\over ,}J_m(\sigma')\}=
{4(n+m)\over g^2}\tanh(\lambda) r\delta '(\sigma-\sigma ')
+[r,J_{n+m}\otimes I-I\otimes J_{n+m}]\delta(\sigma-\sigma')
}
where
$$
\{J_n(\sigma){\buildrel\otimes\over ,}J_m(\sigma')\}=
\{J^a_n(\sigma),J^b_m(\sigma')\}t_a\otimes t_b
$$
here $I$ is the unit matrix
and r, which is given by
\eqn\eqrmat{
r=\half\kappa^{ab} t_a\otimes t_b
}
is the classical $r$-matrix.

\newsec{The beta function}

The beta function for the general sigma model is known \DF. 
It is  given in terms of the Riemann tensor. To first order
\eqn\eqRG{
\mu{\del\over\del\mu}G_{ij}={1\over 4\pi}R_{ij}
}
where $R_{ij}=R^k_{\ ikj}$.

The curvature is given in terms of the Levi-Civita connection
\eqn\eqriem{
R^i_{\ jkl}=
\del_k\Gamma^i_{\ lj}-
\del_l\Gamma^i_{\ kj}+
\Gamma^r_{\ lj}\Gamma^i_{\ kr}-
\Gamma^r_{\ kj}\Gamma^i_{\ lr}
}
where the Levi-Civita connection is given in terms of the metric
\eqn\eqlevic{
\Gamma^i_{\ lj}=\half G^{is}\big(
\del_l G_{sj}+
\del_j G_{sl}-
\del_s G_{lj}\big)
}

The Ricci tensor in our special case was written explicitly by Halpern and
Yamron \ref\HY{M. B. Halpern and J. Yamron, Nucl. Phys B332 (1990) 411.}. We rederive, for 
completeness sake, their result together with the explicit form of the Riemann
tensor which is needed in higher order corrections to the beta function.
Since the metric in our special case  (as well as in \HY) is given by
\eqn\eqGijII{
G_{ij}= L_{ab}e_i^ae_j^b
}
the Levi-Civita connection  has the form
\eqn\eqlc{
\Gamma^{i}_{\ lj}=
\half e^i_a(\del_l e^a_j+\del_j e^a_l )+
\half F^a_{sr}e^i_a(e^s_j e^r_l+e^r_j e^s_l)
}
A tedious but straightforward calculation now gives
\eqn\eqriemann{
R^i_{jkl}=
{1\over 4}U^b_{\ rqs}e^i_b
 e^r_j
e^q_k
e^s_l
}
where
\eqna\eqU
$$
\displaylinesno{
\qquad U^b_{rqs}=
f_{qs}^{\ \ a}f_{ra}^{\ \ b}+
2\big(S^a_{\ rs}f_{aq}^{\ \ b}-
S^a_{\ qr}f_{as}^{\ \ b}+
S^b_{\ qa}f_{rs}^{\ \ a}-
S^b_{\ sa}f_{rq}^{\ \ a}\big)\hfill\cr
\hfill {}+4\big(S^b_{\ ar}f_{qs}^{\ \ a}+
S^a_{\ rs}S_{\ qa}^b-
S^a_{\ qr}S_{\ sa}^b\big)\qquad\qquad &\eqU{}\cr
}
$$
and $ S^a_{\ rs}$ is given in \eqFabc.
The Ricci tensor is now given by contracting indices
\eqn\eqRicci{
R_{ij}=R^k_{\ ikj}=
{1\over 4}U^b_{\ rbs}e^r_i e^s_j
}
where
\eqn\eqUbb{
U^b_{\ rbs}=
(\hat h\kappa_{rs}+
2S^b_{\ ar}f_{bs}^{\ \ a}-
2S^b_{\ sa}f_{rb}^{\ \ a}-
4S^a_{\ br}S_{\ sa}^b)
}
where $\hat h$ is the dual Coxeter number defined by 
$f^{\ \ a}_{br}f_{as}^{\ \ b}=-\hat h\kappa_{rs}$. 
We also  used the fact that $S^b_{\ ba}=0$.  Further simplification can be
achieved using the following identities:
\eqn\eqidenII{
\eqalign{
F^b_{\ ar}f_{bs}^{\ \ a}&=-F^b_{\ as}f_{rb}^{\ \ a}\cr
F^b_{\ ra}f_{bs}^{\ \ a}&=F^b_{\ as}F_{\ rb}^a\cr
F^a_{\ br}F_{\ as}^b&=
f^{\ \ a}_{br}f_{as}^{\ \ b}=-\hat h\kappa_{rs}\cr
}
}
The final result is
\eqn\eqUbbfin{
U^b_{rbs}=2\hat h\kappa_{rs}-2F^b_{\ as}f_{rb}^{\ \ a}
-F^a_{\ rb}F_{\ sa}^b
}
This is in agreement with \HY.
The $\beta $ function equation reads now
\eqn\eqRGfin{
{\del\over\del\log\mu}L_{rs}=\beta(L_{rs})=
{1\over 16\pi}(
2\hat h\kappa_{rs}-2F^b_{\ as}f_{rb}^{\ \ a}-F^a_{\ rb}F_{\ sa}^b)
}
This relation holds whether or not the model is integrable. It is interesting
to see some examples:

$\underline{\hbox{Example 1}}$: The PCM  $L_{ab}={1\over g^2}\kappa_{ab}$. In 
this case
\eqn\eqFFexI{
F^a_{\ rb}F_{\ sa}^b=-\hat h\kappa_{rs}
}
and
\eqn\eqRexI{
R_{ij}={1\over 4}\hat h\kappa_{rs}
e^r_i e^s_j
}
and the renormalization equation is solved simply 
\eqn\eqRGexI{
{1\over g^2(\mu)}=
{1\over 16\pi}\hat h\log({\mu\over\Lambda})
}

where $\Lambda =\exp{(-{16\pi\over {\hat h}g^2(1)})}$.
Clearly the model is integrable along the flow. 

$\underline{\hbox{Example 2}}$: The diagonal anisotropic PCM: 
$L_{ab}=J_a\kappa_{ab}$ (no summation).
we take $SU(2)$ with $J_1=J_2\ne J_3$, and the dual Coxeter number for 
$SU(2)$ is $\hat h=2$.
The Ricci tensor for this case is
\eqn\eqRexII{
R_{ij}=
{1\over 2}(\hat h-\lambda_r(J))\kappa_{rs}
e^r_i e^s_j
}
where
\eqn\eqlami{
\eqalign{
\lambda_1(J)&=
\lambda_2(J)={J_3\over J_1}\cr
\lambda_3(J)&=2-\big({J_3\over J_1}\big)^2\cr
}
}

After changing variables
\eqn\eqnewvar{
t=\log\mu\quad ,\quad x=8\pi J_1\quad ,\quad y=8\pi J_3
}
the renormalization equations read
\eqna\eqRG
$$
\eqalignno{
{\del x\over\del t}&=2-{y\over x}&\eqRG a\cr
{\del y\over\del t}&=({y\over x})^2&\eqRG b\cr
}
$$

Define $z={y\over x}$ then we can separate variables:
\eqn\eqdeI{
x{\del z\over\del x}={2z(z-1)\over 2-z}
}
which is solved by
\eqn\eqdeII{
x^2=a^2{z-1\over z^2}
}
where $a$ is an integration constant. Using the above equations, we 
finally get
\eqn\eqdeIII{
{\del z\over\del t}={2\over a}(z-1)^\half z^2
}
Changing variables one more time $w^2=z-1={J_3-J_1\over J_1}$, this equation
is readily solved by
\eqn\eqsol{
t+const ={a\over 2}A^2\log{w-B\over w+B}-a({B\over A})^2\tan^{-1}{w\over A}
}
where 
\eqn\eqAB{
\eqalign{
A&=\sqrt{\sqrt{2}+1}\cr
B&=\sqrt{\sqrt{2}-1}\cr
}
}
and again it is clear that the model is integrable along the flow.

\newsec{Conclusion}
The main result of this paper is the derivation of the algebraic equations
\eqmaster{}\ as conditions for the integrability of the GPCM. As common in the
study of integrable models, these equations define an overdetermined system.
Naive counting gives $(\dim G)^3$ equations for only $(\dim G)^2$ variables.
It is already remarkable if a solution exists, but this is not enough to
guarantee integrability.
It must also have, as an algebraic variety, 
a dimension greater then zero to allow for a dependence on a spectral parameter.

We showed by
examples that the set of solutions to these conditions is not empty.
These two examples: the PCM and the $SU(2)$ diagonal anisotropic PCM are 
known to
be integrable \KR\PW\ and it is reassuring to see that they actually solve the
system and that the spectral parameter naturally arises. 
However, these are not the only solutions of \eqmaster{}. The search for new
solutions is under current study and will be reported elsewhere.

Another important subject in classical integrable models is the classical 
r-matrix. We note that we were not able to find, in general, 
a set of fields which includes the Lax field and that is closed under 
the Poisson brackets. Another way to attack this problem is to use the 
generalized
Thirring model which gives the same equations of motion.
It has, though, different
conditions for integrability, and
different Poisson structure and it may be easier to identify and to close an 
algebra in this setting\ref\NSII{K. Bardak\c ci, L. M. Bernardo and N. Sochen,
 ``Integrable generalized Thirring model", In preparation.}. 
I believe nevertheless that the example of the PCM
is very revealing and very intriguing. From one hand the appearance of the
two loop affine algebra is surprising since the models manifest only
 $G\times G$
symmetry, on the other hand we must remember that the fields that we defined
are not the conserved currents. Thus one cannot infer, 
that this algebra is a symmetry of the model.

The directions for further research are numerous. In fact we only have scratched
the tip of an iceberg. At the classical level there are
the tasks of adding the antisymmetric tensor and dilaton terms and
deriving integrability conditions for these models. Another direction
may be adding fermions and studying supersymmetric models. There is also 
of course
the problem of finding new solutions to both overdetermined 
differentio-algebraic systems eqs. (3.4),(3.6-7). 
It is also important to find higher conservation
laws, and to find and understand the classical r-matrix. For almost all
of these problems there is a quantum analog.
Yet another direction is to look at the possible relation between these 
models and the
irrational conformal field theories 
(see \ref\H{M. B. Halpern, E.Kiritsis, N.A. Obers and K. Clubok, 
Phys. Rep. 265 (1996) 1,
and hepth/9501144} and references therein). We hope to address some of these
questions in the future.
\vskip3.5cm
\centerline{\bf Acknowledgment}
\vskip0.5cm
I would like to thank Prof. N. Yu. Reshetikhin who introduced to me this 
problem
and shared with me his insights, and Prof. K. Bardak\c ci for his constant
intertest, encouragement and numerous discussions.
I also benefited from discussions and remarks of
L. M. Bernardo, J. D. Cohen and M. B. Halpern.

\acknowledge\
\listrefs
\bye